# Business and social evaluation of denial of service attacks in view of scaling economic counter-measures

L-F Pau, Prof. Mobile business Copenhagen business school, and Rotterdam school of management lfp.inf@cbs.dk

ABSTRACT: This paper gives an analytical method to determine the economic and indirect implications of denial of service and distributed denial of service attacks. It is based on time preference dynamics applied to the monetary mass for the restoration of capabilities, on long term investments to rebuild capabilities, and of the usability level of the capabilities after an attack. A simple illustrative example is provided for a denial of service on a corporate data centre. The needed data collection methodologies are categorized by classes of targets. The use of the method is explained in the context of legal or policy driven dissuasive, retaliation or compensation/ restoration actions. A concrete set of deployment cases in the communications service and transport industries is discussed. The conclusion includes policy recommendations as well as information exchange requirements.

INTRODUCTION

This work in progress aims at addressing two strategic aspects of cyber-warfare mostly via communications networks and IT applications: a) first to take a total economic and social view in the assessment of evaluating damages of a cyber-warfare attacks on a society or business target; b) scaling a trade, economic, or legal retaliation or dissuasion for decision makers. It is assumed that the target of the attack does not in general have itself any or sufficient defence or attack means, so that a corporate or national level may decide ex-ante (dissuasion) or ex-post (retaliation, compensation) to scale a business defence affecting the economic sphere of the attacker. Such an approach is also relevant sometimes when attacker cannot be identified and localized precisely, so that the economic sphere of the attacker is restricted to business networks to which the attacker belongs.

Traditionally the damage assessment has been considered "binary" and limited in time, in that the target was considered to be rendered totally dysfunctional until full restoration only of its information and communication capabilities. Lessons learnt tell us that other organizational, physical, human and social capabilities are to be counted as representing often larger collateral damage of the attacks; their restoration eventually takes quite some time, especially if the surrounding society does not have enough civil defence means and skills in place. Vice-versa, sometimes, the replacements made to infrastructure damaged by the attack will be less obsolete leading to better future robustness. To address this issue, the approach is to capitalize on the ability of cost-benefit analysis to bundle into the internal rate of return both tangible and some intangible effects .The internal rate of return expresses the time preference on tangible and intangible assets ,old and new, which gives a break even net present value over the long term. It is then proposed to treat short term dynamics of this internal rate of return , when exposed to a Brownian shock linked to an attack affecting the command and control node for the society or business target which have their normal long term equilibrium return rates.

Assuming the dynamic time preference resulting from a cyber-attack, it becomes possible to estimate all of the following :a) the incremental monetary mass needed short term for restoration of equilibrium business and social capabilities; b) long term investment over a given pay-back horizon needed over time to restore and improve capabilities to get back to the equilibrium rate; c) the value of the assets degraded by the cyber-attack as short term and long term restoration measures impact the target.

Apart from relevance in a national or corporate budgeting process, such a three-dimensional scaling of compensation, retaliation or dissuasion gives decision makers a way to communicate efficiently around them and to implement such counter measures against the attacker's economic sphere while referring eventually to a game theoretical equilibrium required by legal/treaty provisions.

As a conclusion, the proposed methodology empowers decision makers to scale eventual economic counter-measures or threats against attackers, the efficiency of which cannot be guaranteed as economic-social effects may not always impact attackers but surely their surroundings, and as the resolution of decision makers may also vary. It will be up to the reader to assess relevance in her/his own context, while this project has assessed some concrete cases. This project has also been motivated by specific concerns and abilities of wireless communications operators.

1. SURVEY

The cyber attacks considered in this paper (denial of service DOS , and distributed denial of service DDOS) are those damaging information, capabilities, and sometimes network and infrastructure elements owned or operated by a target, with resulting damages not only to the target but also to third parties dependent on this information, or those networks and infrastructure [1, 14, 16]. Damage assessment is considered difficult, as the intrusions and attacks cannot always be detected short term [2, 15, 17]. Nevertheless, large economic and social impact is felt, reaching from a unit in an organization to whole sectors; have been carried out as part of earlier work: descriptive assessments of the impact from surveys with input-output analysis of effect from outages and propagation models (e.g. [3, 7]), evaluations of incentives and investments to protect the information infrastructure (e.g. [4, 6, 8]), and evaluations of cyber-insurance premiums in relation to security procedures [5, 20]. Very few papers deal with models for damage assessment , which would allow a company to qualitatively and quantitatively estimate possible financial losses due to partial or complete interruption of connectivity ; in [9] a systems engineering approach is taken, while in the present one an economic and business approach is taken and a simple numerical example is given in Section 4.

Also we will address in Section 5 the use of damage assessment estimates on legal grounds for retaliation or compensation [18, 19]. A distributed Denial of Service attack aims to deprive legitimate users of a resource or service provided by a system, by overloading the system with a flood of data packets, thus preventing it from processing legitimate requests. Therefore it is necessary as in [10] to take into account the doctrines governing the allocation of liability among key players in a distributed denial of service attack. Such doctrines are well established and based on common law tort principles and policy considerations.

Regarding related types of attack, such as malware, viruses, identity theft, exploiting vulnerabilities in control software / management functions/ protocols (such as DNS and BGP errors, lack of authentification of users, services or flows, payment solutions vulnerabilities), some studies like those of Ferris Research and Gartner Research have shown the huge business impact thereof as well as the very high handling plus restoration costs. But such estimates are at best interview based, and lack an analytical framework.

## 2. THEORETICAL BASIS

*Comparison with economic theory*

One way of looking at the economic consequences of a denial-of-service is to consider that the target has a diversity of assets included in a portfolio , each with varying life-cycles, and that any attack affects the overall value and sustainability of the portfolio. Whereas in economics and finance the typical research question is one of asset allocation in view of returning some performance goals [13], the cyber-warfare economics question is one of asset preservation over time. Another difference with economics and finance is that in these fields' risks and returns are usually mutualised across populations of owners or users via legal contracts, in cyber-warfare economics the target normally stands alone at the time of attack with all risks and must have made all required preventive investments. The only subfield of economics where some common features can be identified, is the area of pension economics where the retired person wants to maintain over time a purchasing power level, although here again assets are a mix of own assets and mutualised assets.

Regarding the definition of capabilities exposed to an attack, they are defined at any time as the net difference between a normal time-dependent operational capability profile of the attacked entity, and the complete or partial combined effect of the attack and of restoration measures on normal capabilities due to the nature of the attack and restoration processes. Consequently, dynamics play an important role, and the proposed methodology encompasses situations with a net reduction in capabilities. If the attack on one target involves reduced capabilities of other asset owners (like in the case of a "netbot", or the halving of transmission rate capability by the TCP protocol in case of a transmission error / congestion) one can either take a systemic view or the view of the target alone.

Regarding the description of the stochastics of attack processes, only attack specific process specifications with related methods would allow to model them closely, but macro-level approximations by known or tailored distributions already provide a good basis.

Regarding the restoration process, it is also to have its specific dynamics. However, restoration is supposed to be possible, at a cost, but not impossible, thus implying that data protection, integrity and security must be in place. In the case of data loss prevention (DLP) , the Ponemon Institute has estimated from commercial cases the cost of data loss to 100 k Euro- 5000 k Euro of which 36 % due to commercial losses and lost customers, and 36 % from loss of portable data storage. Although VOIP content is vulnerable, repeated calls remain possible in general.

*Proposed methodology: time preference dynamics*

The proposed methodology is to assume that the target applies different time preferences to the assets in its portfolio, where the time preference profiles express the urgency at which restoration of capabilities must be carried out in view of a time distributed attack (including a shock) degrading suddenly specific assets in the portfolio. In economics, **time preference** (or "discounting") pertains to how large a premium a user will place on usage nearer in time over more remote usage.

Taking one class of assets, assume that the time preference rate r(t) fluctuates around an equilibrium level r(eq) while subject to a Brownian point process W(t) .The short term dynamics are modelled by [12]:

dr(t) = a (r(eq)-r(t))dt – V.dW(t)  (1)

where :
- r(t) is the short term time spot preference at time t for a given asset,  0<r(t)<1
- a is the intensity of the feedback force towards the equilibrium time preference r(eq)
- r(eq) is the equilibrium time preference for that given asset
- V is the volatility of the time preference fluctuations
- W(t) is the stochastic Brownian point process driving the attack diffusion process

*Monetary mass requirements for restoration of capabilities*

The incremental monetary mass dM (t) needed short term for restoration of equilibrium capabilities of the asset can then be determined. Assuming for simplification purposes the short term time preference rate r (t) to drive short term interest rate dynamics by near a constant rM:

dM(t) = M(t-dt). (r(t)+rM).dt    M(0)=M0

where :
- M(t) is the monetary mass used short term to invest in rebuilding the asset capability to its levels just before t=0 where monetary mass represented by the asset value was M0
- rM is the fixed increment to the short term time preference producing the short term interest rate payable to finance the rebuild of the capabilities

*Long term investments to rebuild capabilities*

The long term investment K (t) over a given horizon TK needed long term to restore and improve the assets capabilities to get back to the equilibrium time preference rate r (eq) can be expressed as follows:

dK(t) = K(t) [r(t).dt + V.BetaK.( dW(t)+Lambda.dt) ]
BetaK = (1-exp(-a.TK))/a

where :
- K(t) is the long term bond-like investment needed over the horizon TK to restore asset's capabilities

- K(0) is the initial annuity value of the assets capability value over the horizon TK
- TK is the time horizon to rebuild and possibly improve on the asset's capabilities; this parameter is essential in all practical cases
- Lambda is the premium by unit risk needed by the market to support the randomness over the real time preference
- BetaK s a constant

*Usability of the capabilities over time after an attack*

The value of the assets degraded by the cyber-attack as short term and long term restoration measures impact the target, is linked to a specific usability risk characterization WA (t) of the asset's capabilities. The change in the degree of usability A (t) of this asset, bounded between 0 and 1 is:

dA(t) = A(t). [r(t).dt +V (dW(t)+Lambda.dt) +VA (dWA(t)+LambdA.dt)]

where:
- the first term in the parenthesis is the effect of short term restoration via the monetary mass investment
- the second term in the parenthesis is the contribution from long term fixed horizon asset capability rebuild
- the third term is the reduction in recovery speed linked to the volatility and risk in the asset's specific capabilities as they impact its degree of use
- A(t) is the effective degree of usability of the asset , A(eq)=1
- VA is the volatility of the asset's capabilities usability risk
- WA(t) is the Brownian motion of the usability risk characterization of the asset's capabilities
- LambdA is the premium by time unit in unit usability risk needed by asset users to support the randomness over the asset's usability risk.

The unique property of this model is that all time preference variations are subject to the short term time preference and that the risk exposure, which is here the investment needed to restore the asset's capabilities, is by one bond-like financing the duration of which determines the size. The usability of the asset is a Brownian movement correlated with the time preference rates over time. Another characteristic of this model is that it is decoupled from the initial asset valuation , which can be tailored to specific cases and rely on data pre-existing to an attack (see Sections 4 and 5).

## 3. A SIMPLE NUMERICAL EXAMPLE

*Scope*

This very simple example does not allow to show and exploit all the dynamic effects taking place, but to show how a concrete situation can lead to estimations of short term and long term financing needs tied to the time preference expressed. It also shows that, even if financial means are made available to rebuild capabilities, the actual restoration time of usable capabilities is very much subject to the stochastic distribution properties and to the quality of actual means for capabilities restoration. It also leads in Section 4 to further data collection methodology considerations.

*Description*

The numerical example pertains to a data centre in a company, with a scrap value of 10 MEuros, running services to support company operations. The equilibrium state is one where all services operate 100 % to support all divisions and operations with a company turnover of 500 MEuros/year; furthermore client capabilities are dependent on the company's operations being supplied to them for another 500 MEuros /year (treated as contingent liabilities). The equilibrium time preference r (eq) is equivalent to the company's net operational profit margin from operations r (eq) = 50 %/year, approximated as 0.5/ (365x24) = 5.7E-05 /hour. The short term monetary interest rates are only about 10 %/year, so that rM= -4.76E-05/hour. A full instantaneous attack W (0) =1 on the data centre at time t=0 reduces services usability to A (0) =0 with a minimum nominal restoration time of TK= 3 months for all resulting services and operations to internal divisions and third parties after such a disruption. The attack lasts dt= 1 hour , taken also as time increment, creating a shift in the time preference to a very high spot time preference value ; the maximum which can be chosen is r(1 hour)=1, meaning the target wants perceptually all measures to be taken for immediate recovery of the data centre . With a maximum volatility in time preference fluctuations of V=1 /hour, the needed reactivity becomes: a ~ 5.8E-05.  Post attack, the short term time preference grows tremendously leading to a strong rise in perceived short term monetary flows for restoration dM(1) of  slightly under 10 MEuros/hour ; this expresses the perception that the data centre must be restored at once . The total capability value of the assets over TK=3 months is 250 MEuros with an hourly annuity of 115 740 Euros. With a risk premium Lambda= 0.2, the initial long term investments dK (1) needed to recoup lost supplies to customers, and to rebuild the capability, can be estimated at about 235 M Euros.  For the usability risk WA (t) a simplified linear decreasing profile can be taken over the restoration period TK, that is WA (t) =1-(t/TK); we also assume LambdA=0. However,  the quality and efficiency of the restoration are highly volatile especially in downstream supply chains from the company ;  this leads to the  usability of the target's capabilities only increasing again (dA(t)/dt >0) , despite a high time preference,  if the volatility VA is less than 1,2*TK .  Half of the overall capability is only restored at time 0,5/ (1,2-VA/TK) which can be longer than TK= 3 months for some values of VA.

4. APPLICATION AREAS AND DATA COLLECTION METHODOLOGIES

This paper cannot give cases or fictive examples for all the application areas for which economic and social impact of denial of service need to be quantified. This Section only serves to survey such areas by categories and to give when known established approaches to assess relevant data to be fed into the calculations.

1. *Public services*

The denial of service of public services on a national basis or on an agency basis (administrative services, social services, water, air traffic, waste management, financial payments), have wide ranging consequences where the indirect impact encompasses prejudice caused to citizens (in their ability to act, to get benefits or to contribute tax etc) measured in time lost, benefits / contributions lost, and of qualitative damage (health, safety, administrative registrations etc). In this field, traditional cost-benefit analysis of tangible and intangible services applies. As to the setting of the time preference rates, they should be high for those public services where public authorities by law have obligations of service continuity, while they would be less and derived from minimal service obligations in other cases.

2. *Company products and services*

In this case, the applicable methodology to the data collection is the one used for corporate liability insurance assessment. This includes loss of capabilities (physical, raw material and service related) with their replacement, loss of revenue due to non delivery in time, physical loss of output such as manufacturing with associated logistic and CRM overheads , indemnification of human resources if work or life is jeopardized, and indirect loss and damage to clients. As to the setting of time preference rates, in-company rates should correspond to the average return on assets or operational margin (whichever is largest) within the sector in which the company was denied services, while the same would apply for the clients in their respective sectors.

3. *Loss of shared infrastructure*

There is no established methodology to cover loss of shared infrastructure, "critical" or not (such as communication or transportation networks, denial of service of a satellite by jamming, etc). However the normal approach would be to make the inventory of the lost capabilities (physical and service related) by infrastructure operator, of lost revenues by infrastructure operator including claims payable to customers under contract terms, of verifiable loss and damage by individual and institutional users, and moreover of social costs to the same. As to the setting of time preference rates, this is a difficult issue as infrastructure suppliers quite often do not have contractual quality of service obligations. On the contrary, suppliers of "critical" infrastructure whose control systems may have been compromised, bear a responsibility beyond just service provisioning, and there recovery processes may be longer. Judgment would have to be applied to the time preference of the infrastructure operator (normally very high but not coupled to financial rates of returns) and to the users taking diversity into account. For users the principle of setting the time preference could be based on the tolerable postponement of the access and use of the shared infrastructure to next normal period (such as shift by e.g. one day, or to next available equivalent infrastructure provider).

4. *Technology providers*

Some well known technology providers in such areas as communications, software, control systems, transport technologies, biomedical devices, etc.., may be liable to claims by their customers for vulnerabilities in their products, although third parties are those exploiting them. While the "customer cum users" would know the attack profiles, while not always knowing the technical roots for the vulnerabilities, technology providers may benefit from the proposed framework for risk assessment if they share attack profiles with their customers. The risk assessment method in turn allows them to quantify reasonable levels of investments in improving the technologies and their distribution mechanisms.

## 5. DENIAL OF SERVICE IMPACT ANALYSIS USAGE PROCESS

The concept is to use the damage assessment methodology of Section 2 , with its different time scales, to specific data collected by established methodologies moderated by neutral judgement (like best practices or eventually arbitration courts) (see Section 4) , to calculate estimates of the set of damages . Such assessments must be transparent and done by neutral parties.

The assessed damages can then be used by executive authorities for a spectrum of actions:

- Dissuasive process: preemptively to a denial of service, by policy makers or companies, to announce that these claims would be raised if an attack occurs. The policy makers or companies may not have evidence yet or from past cases to identify the attackers, but may communicate to make such a categorization of attackers credible and visible to attackers .Also, subject to proper later judicial tracing and identification of the attackers, the policy makers or companies would communicate that they intend to recover the amounts of the claims by all legal means in case of an attack. As the average cost to attackers of a cyber-attack is usually small, dissuasion followed by retaliation or recovery may be of some concern to attackers or their backers.
- Retaliation process: if the attackers are traced and identified by technical and/or judicial means, or if strong assumptions and partial evidence exist (e.g. from IP addresses, software code structure, software forensics, etc…), legal or forceful retaliation would be done for the same size of claims against direct or indirect interests of the attackers. One obvious instance of this would be to seize quarantine or destroy the physical and communications assets used by the attackers, or assets owned controlled by them. This may happen in a judicial framework (with fines and penal measures) or an international treaty framework, but may be replaced by policy maker coercitive decisions including offensive means.
- Compensation / Recovery process: if the attackers are traced and identified by judicial means, and can be put on trial, this process would use the damage assessments as normally done in a judicial court procedure. In this case however the data collection methodology and data would be subject to a contradictory evaluation, there may be issues of sovereignty leading to inability of enforcement/ extradition, and the delays involved are normally quite long.
- "Keep silent" process: There is of course a fourth process, which is to ignore attacks, keep silent, report nothing, and not to sue, often for "image» reasons. It is unfortunately very common that banks, communications and infrastructure operators so far do not report attacks and even figure out other reasons vis-à-vis their users.

It is conjectured that the main practical relevance of the proposed method is for dissuasive and retaliation processes, resting ultimately on the ability of the asset owner / target to carry out and update his own exposure valuations based on estimates related to user and client damages (tangible and intangible).

This same conjecture is obviously reinforced by the consideration that the tracing and identification of the true attackers may not always be possible, or may take so much time, that the strategy to use a recovery process may not work while a dissuasive or retaliation process may have effects when used together.

Likewise, if attackers are using innocent identifiable resources, a recovery process would take time establishing that they are not responsible, while giving time to the responsible attackers.

It should not be forgotten that cyber-attacks against corporate assets often are initiated from inside the company or past employees, which too opens up for a combination of dissuasive, retaliation and partial recovery processes.

Finally, as some types of defensive measures (such as anti-virus) have fast deployable get-around's known to attackers, dissuasion and retaliation processes may in some cases be the only way forward.

6. APPLICATION CASES

This research has found its way into a number of deployment cases summarized below spanning all categories identified in Section 4:

- Public services

  Case: minimal public transport service under employee strikes (Western Europe)
  Contribution: the proposed method allowed to determine the public damage- number of employees on strike curve, allowing for the union and the employer to settle on a minimum service level.

- Company products and services

  Case: corporate liability insurance estimate for a Scandinavian CRM provider
  Contribution: the customer relationship management (CRM) company's services were outsourced by several operators in the communications and credit card fields. The contracts between these operators and the CRM service supplier stipulated damage claims should the CRM supplier not be operational. The method allowed the CRM provider to determine the liability insurance amount it had to get cover for vis-à-vis cyber attacks to compensate its customers.

- Loss of shared communications infrastructure

  Case: attack on 3G operator BSC with partial recovery via other operator(s)
  Contribution: The wireless 2G and 3G base system controller manages the connectivity to and from radio base systems (RBS). Due to bad network management or practices, some BSC are not totally immune from certain types of attacks. When redundancy and restoration procedures have failed, radio coverage and connectivity may be lost unless back-up is activated from other operator's BSC (when feasible). Such operators have to be compensated, as well as possibly some wireless service users under contractual terms, and total damage assessment with/without insurance may be necessary.

Mobile networks not only provide great benefits to their users but they also introduce inherent security issues. With respect to security, the emerging risks of denial of service (DOS and DDOS) attacks will evolve into a critical danger as the availability of mobile networks becomes more and more important for the modern information society. There are ways to mitigate the attacks by adding minimal authentication to the radio channel assignment protocol, but this too has business implications and requires risk assessment. At the same time, via subscriber management, interoperable management and signalling / control networks, they carry the potential for tracing and retaliation measures, besides lawful interception in support of legal procedures. In particular is highlighted the retaliation process which international inter-carrier settlements allow for, as such agreements reach out worldwide.

Finally, it has been brought to the attention of the author, that other applications exist, e.g. in the case of water distribution protection, where attacks have wide reaching implications, and where physical-chemical forensic evidence may be collected. In this case the attack has both time-based as well as spatial distributions.

CONCLUSION

While law and jurisprudence regarding denial of service and other cyber-attacks is making slow progress in both national and international arenas, this paper presents a quantitative approach respecting attack and restoration dynamics likely to be used in dissuasion as well as in retaliatory processes, in the hope that ultimately attackers will feel a largely missing retroaction. It may also allow institutions and companies to determine by self-analysis in the presence of a given threat profile, which assets to protect in priority on economic, business and social grounds.

In the event international organizations like GATT, European Space Policy Institute, OECD or the European Parliament ("Declaration on the reinforcement of international security", 25 March 2003 and report to the Council of 11 December 2008) also embark on putting an economic and social measure to cyber-attacks, supplemented by constraining legal measures, instead stating of political / cultural or defence values only, this research may give elements of the analysis.

Specific policy recommendations linked to the above research and the deployed cases, would be the following:
-in international commercial contract law, allow for compensation and information exchange clauses whereby attackers using one party's facilities or services to mount an attack on the other party, may retaliate against the attackers on the basis of damage assessment and evidence provided by the other party; an example of this are international communications operators inter-operator settlement procedures;
-enhance auditing procedures, to verify the basis for insurance or damage claims in the case of cyber-attacks;
-mandate reporting and information exchange about attacks to designated governmental bodies, for sharing of attack profiles and partial evidence (like envisaged by the EU).

Just as technical vulnerability reduction demands collaborative efforts between users, technology providers and operators, the business and social impact assessments also demand such collaboration and information exchange, besides internal due diligence. The issue is which governments, players and sectors, like the communications industry, will take concrete steps in this direction. One reason why this is an issue is that "patches" and additional costly imperfect technologies are too often preferred to demanding and longer lasting technical, legal, architectural and economic measures. It is in this context that humanities, economic and social disciplines can clarify the way towards peace in cyber conflicts [21].

What this research does not allow to do is to account for interdependencies between targets and attackers, or proxies to the attackers, due to cross-ownership, exclusive agreements, shared infrastructure (buildings, communications, transport, and energy), geo-economics and political / cultural / social influence.